# DISCRETE THERMODYNAMICS OF LASERS

B. Zilbergleyt,
System Dynamics Research Foundation, Chicago, USA
sdrf@ameritech.net

## INTRODUCTION

Lasers are transforming input energy flow into the output; as energy transformers they should obey the thermodynamic laws. As strange as it is, the only found really thermodynamic approach was Einstein's prohibition on the ability of a 2-level laser to function [1], based on Boltzmann statistics and on the equal pump up – drop down probabilities for the excitable atoms. Einstein considered a light emitting process in isolated system with no pumping energy from outside; needless to say that a real lasing device is an open system and lives different life. Another found relevant publication [2] is currently available with reasonable efforts only as abstract. The bulk of publications that investigated energy dissipation and consequential heating of the laser devices are not relevant to the topic of this paper.

The objective of this paper is to outline a non-classical, discrete thermodynamic model of lasers as energy transforming open systems. Discreteness in this case means usage of finite differences instead of derivatives, which leads to important advantages in complex systems treatment as it has been shown for chemical systems [3]. In case of lasers, such an approach can describe population dynamics and particles distribution between levels, and eventually create a thermodynamic theory of lasers.

Initially the idea of discrete thermodynamics applications to lasers was released by the author in [4]. In the present paper derivation of the basic expression, some thoughts and conclusions are changed towards - in our opinion - more adequate description of the lasing devices and in a manner that is better to understand by the laser professionals.

## THERMODYNAMIC CONSIDERATIONS

The ideal laser system allows to pump as much as possible amount of excited atoms from their ground state (subsystem A) to the upper level (subsystem A*), then holds them over there as long as necessary until the excited atoms get back to the ground state due to the light emission.

Suppose that laser contains 1 mole of the excitable atoms. At zero pumping force the laser stands in internal thermodynamic equilibrium (TdE), i.e. thermodynamic equilibrium between the subsystems A and A*. Their equilibrium populations are $\eta$ ($<1$) moles of the A atoms and $(1-\eta)$ moles of the A* atoms; the ratio between them is defined by the Boltzmann distribution

(1) $\qquad (1-\eta)/\eta = \exp(-h\nu/kT)$.

In open laser system the upper level population increases due to the energy pumping and absorption

(2) $\qquad A + \varepsilon_{in} \rightarrow A^*$,

while stimulated or spontaneous discharge of the accumulated energy via light emission

(3) $\qquad A^* \rightarrow A + h\nu$

returns the excited laser population back to the ground.

Neglecting the heating of laser body x due to energy dissipation that may be prevented by cooling, lasers work at permanent pressure and temperature. Characteristic function in this case is Gibbs' free energy. In the laser system its change for the A→A* transition is

(4) $\qquad \Delta G = \Delta G^0 + RT\ln[(1-\eta)/\eta]$.

At TdE, in absence of the pumping force $\Delta G = 0$ combination of (1) and (4) leads to

(5) $\qquad (1-\eta)/\eta = \exp(-\Delta G^0/RT)$.

Multiplying both tiers of the exponent power in (1) by the Avogadro number $N_A$ one can get expression for the standard change of Gibbs' free energy value, reduced by RT

(6) $\qquad \Delta g^0 = h\nu N_A/RT$,

or, after gathering the universal constants into $\lambda_v = hN_A/R$
(7) $\quad\quad\quad\quad\quad\quad\quad\quad\quad\quad\quad\quad\quad\quad\quad\quad\quad\quad\quad\quad\quad\quad\quad\quad\quad\quad\quad\quad\quad\quad\quad\quad \Delta g^0 = \lambda_v \nu/T,$

$\lambda_v = 47.99$ if the frequency is expressed in terahertz. Now the amount of the ground level atoms at TdE is
(8) $\quad\quad\quad\quad\quad\quad\quad\quad\quad\quad\quad\quad\quad\quad\quad\quad\quad\quad\quad\quad\quad\quad\quad\quad\quad\quad\quad\quad\quad\quad\quad\quad \eta = 1/[1+\exp(-\lambda_v \nu/T)].$

For the most important part of the light spectrum, from far infrared to short wave UV, the $\Delta g^0$ values vary from $-63.040$ for to $-246.720$ dimensionless units at T=300K. It is well known that at TdE the $\eta$ values are very close to unity, i.e. the huge majority of the excitable laser dwellers are sitting on the ground as it takes place in real devices. On the other hand, that makes all frequencies within the above IR-UV range thermodynamically undistinguishable.

**THERMODYNAMIC EQUATION OF STATE OF THE LASER SYSTEM**

If x moles of atoms were pumped off the ground level up, the laser system state proximity to TdE may be characterized by the system state dimensionless coordinate
(9) $\quad\quad\quad\quad\quad\quad\quad\quad\quad\quad\quad\quad\quad\quad\quad\quad\quad\quad\quad\quad\quad\quad\quad\quad\quad\quad\quad\quad\quad\quad\quad\quad\quad\quad \Delta_e = (\eta - x)/\eta.$

Obviously at TdE x=0 and $\Delta_e = 1$. We define the shift from TdE as $\delta_e = 1 - \Delta_e$, or
(10) $\quad\quad\quad\quad\quad\quad\quad\quad\quad\quad\quad\quad\quad\quad\quad\quad\quad\quad\quad\quad\quad\quad\quad\quad\quad\quad\quad\quad\quad\quad\quad\quad\quad\quad\quad\quad\quad \delta_e = x/\eta.$

At TdE $\Delta_e = 1$, and $\delta_e = 0$. When the whole population is pumped up, $\Delta_e$ and $\delta_e$ exchange their values.

The laser system is driven by the pumping force out of TdE until a new stationary state is achieved. From basic definition of equilibrium as a balance of acting against the system internal and external thermodynamic forces (TdF) [5]. The balance is given by Onsager equation [6]
(11) $\quad\quad\quad\quad\quad\quad\quad\quad\quad\quad\quad\quad\quad\quad\quad\quad\quad\quad\quad\quad\quad\quad\quad\quad\quad\quad\quad\quad\quad\quad\quad\quad\quad A_i + a_{ij}P_j = 0,$

the first term is well known thermodynamic affinity, which in discrete version is $A_i = -\Delta g/\Delta_e$ [7]. The second term contains reduced Onsager coefficient $a_{ij} = a_{ij}/a_{ii}$ and external, pumping thermodynamic force that equals to $P_j = -E_p/(RT)$, where $E_p$ is the energy absorbed by 1 mole of the excitable atoms. In our case T=300K and RT=2.44 kJ/mol. Taking into account that relation, after through multiplication by $\Delta_e$ and substitution $a_{ij} = a_{ij}/RT$ we turn equation (11) to
(12) $\quad\quad\quad\quad\quad\quad\quad\quad\quad\quad\quad\quad\quad\quad\quad\quad\quad\quad\quad\quad\quad\quad\quad\quad\quad\quad\quad\quad\quad\quad \Delta g - a_{ij}(1-\delta_e)E_p = 0.$

To obtain the laser system equation of state from this precursor, we have to eliminate external factors leaving only variables and internal parameters. Let us express the pumping energy via the laser system reaction to pumping as a power series of the system shift from TdE, which in a simple case may look like
(13) $\quad\quad\quad\quad\quad\quad\quad\quad\quad\quad\quad\quad\quad\quad\quad\quad\quad\quad\quad\quad\quad\quad\quad\quad\quad\quad\quad\quad\quad\quad\quad\quad\quad E_p = \Sigma_{0 \leq k \leq q} w_k \delta_e^k,$

q is a loosely defined system complexity parameter. In case of laser system it may be identified roughly with the number of possible transitions between the levels, i.e. two in case of 2-level equilibrium (one up and one down). Let's move $w_0$ out of the sum in (13); new weights $w_{k0} = w_k/w_0$ are still unknown and we have to set them to unities. It's very easy to see that after placing expression (13) for $E_p$ into (12) and after multiplication $(1-\delta_e)$ by $(1+\delta_e^1+\delta_e^2+\ldots+\delta_e^q)$, all powers of $\delta_e$ but q annihilate, and (12) turns to the final equation of state for the open laser system as a logistic map [8]
(14) $\quad\quad\quad\quad\quad\quad\quad\quad\quad\quad\quad\quad\quad\quad\quad\quad\quad\quad\quad\quad\quad\quad\quad\quad\quad\quad\quad\quad\quad\quad \Delta g - \alpha(1-\delta_e^q) = 0,$

where new coefficient is $\alpha = w_0 a_{ij}$. This laconic logistic map may be extended using (5) and taking into account the population numbers, $\eta(1-\delta_e)$ on the ground and $[1-\eta(1-\delta_e)]$ on the upper level
(15) $\quad\quad\quad\quad\quad\quad\quad\quad\quad\quad\quad\quad\quad -\ln[(1-\eta)/\eta] + \ln\{[1-\eta(1-\delta_e)]/\eta(1-\delta_e)\} - \alpha(1-\delta_e^q) = 0.$

Map (15) contains one variable $\delta_e$ and two parameters of the new theory, the above introduced q and $\alpha$; the latter is known as the growth parameter in the theory of populations [9]. In our approach it defines the rate of the laser system shift from TdE by the pumping force. Before proceeding with solutions, we'd like to mention one interesting feature of this map. It was already mentioned that $\eta \approx 1$; such approximation in (15) in combination with (1) gives us a variation of the basic map
(16) $\quad\quad\quad\quad\quad\quad\quad\quad\quad\quad\quad\quad\quad\quad\quad\quad\quad\quad\quad\quad\quad\quad\quad\quad \lambda_v \nu/T + \ln[\delta_e/(1-\delta_e)] - \alpha(1-\delta_e^q) = 0.$

Let's neglect the A*−population at TdE ($1-\eta \approx 0$!); then, if x moles of A moved to A* the population ratio is $\pi = x/(\eta-x)$; substituting expression for x from (10) we get $\pi = \delta\eta/(1-\delta\eta)$, or $\pi = \delta/(1-\delta)$. That changes expression (16) to more explicit

(17) $$\lambda_\nu \nu/T + \ln\pi - \alpha(1-\delta_e^q) = 0.$$

All logistic maps are solved usually via iterations.

## SOLUTIONS TO THE LASER LOGISTIC MAP

We simulated open equilibria of the 2-level lasers, simulation results were graphically interpreted in Fig.1-Fig.3 using *dynamic inverse bifurcation diagrams* (DIBD, see [3]), that is shift $\delta_e$ vs. pumping force $P_j$. Both values are dimensionless.

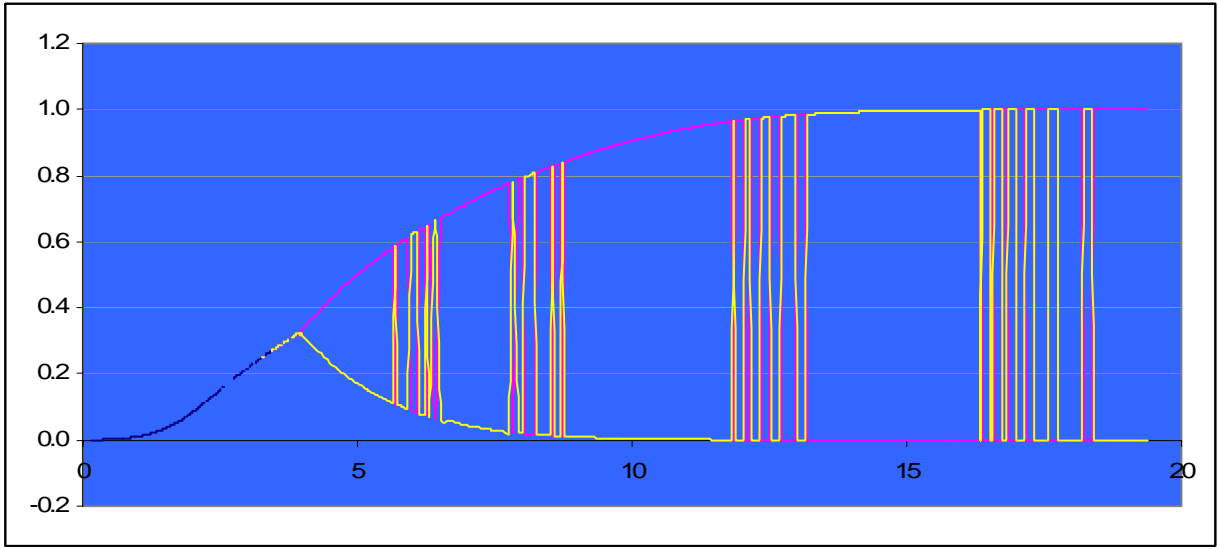

Fig.1. DIBD for the 2-level laser system at q=2.

As the pumping force and amount of activated atoms increase, the system equilibrium shifts from TdE following the thermodynamic branch. The branch looses stability at certain value of the force, splits by two bifurcation branches, and we encounter a bi-stability [10]. Vertical lines mean instability of bifurcation lines; in real laser, downward lines on Fig.1 correspond to spontaneous transitions, i.e. spontaneous emission.

Fig.2 shows short wave UV dynamic diagram, stretched horizontally. All the transitions frequency spectra (do not confuse with spectra and frequencies of the emitted light!), constituted by the immediately following each other up and down transitions are line spectra within the studied light frequency range. The ascending transitions are driven by external pumping power, bringing a part of the laser population to the upper level. Being pumped up, the excited atoms cannot accumulate more energy from the pumping flow: a stress, caused by the pumping "pressure", forces the system to discharge itself by spontaneous or assisted irradiation of light along the descending line. This result represents an ***independent proof of the Einstein's prohibition on the 2-level laser, based exclusively on discrete thermodynamics and extended to open laser systems***.

It was found impossible in this research to get rid of the lines on the pumping force-system shift diagrams by changing the simulation parameters - this linearity seems to be imminent to the model of laser, offered by this theory as well as spontaneous emission in lasers at all. At the accepted simulation parameters, the first linearity starts at $\delta_e \approx 0.6$. If $\eta \approx 1$ that leads to $\pi \approx 0.45$, the value that is more than two times lower than population inversion threshold. Results of this paper prompt us to state that it is definitely very costly and ineffective to create and maintain population inversion in 2–level laser, as it is usually thought [11]. More light might be shed onto instabilities of bifurcation branches by using lower values of η in simulation. Indeed, at η≈1 a very strong force returns any excited atom back to the ground. The less is the η value the easier is to achieve the same shift and the weaker is the returning force, and the less instability should be expected. Unfortunately, the TdE ground population unambiguously depends upon the frequency of emission, and cannot be changed if we need a certain wave length of the emitted light.

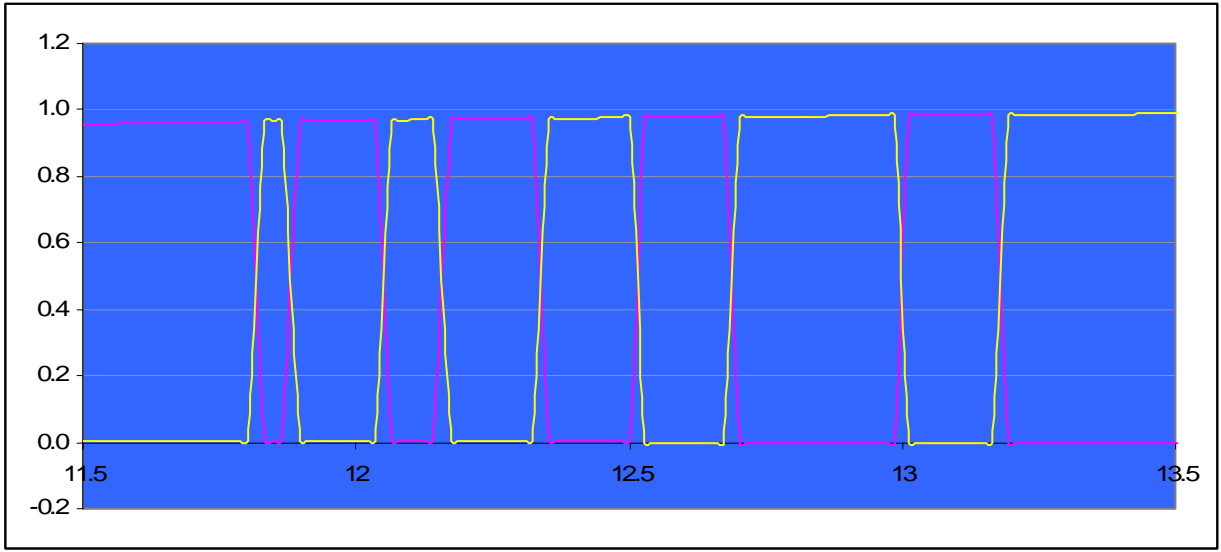

Fig.2. A part of the DIBD from Fig.1, stretched to give a clear show that bifurcation instabilities are synchronous. Spontaneous emission occurs along the lines that go down.

## 3-LEVEL LASER

In real lasers this obstacle is bypassed by adding additional levels, the third and the forth. Here we will

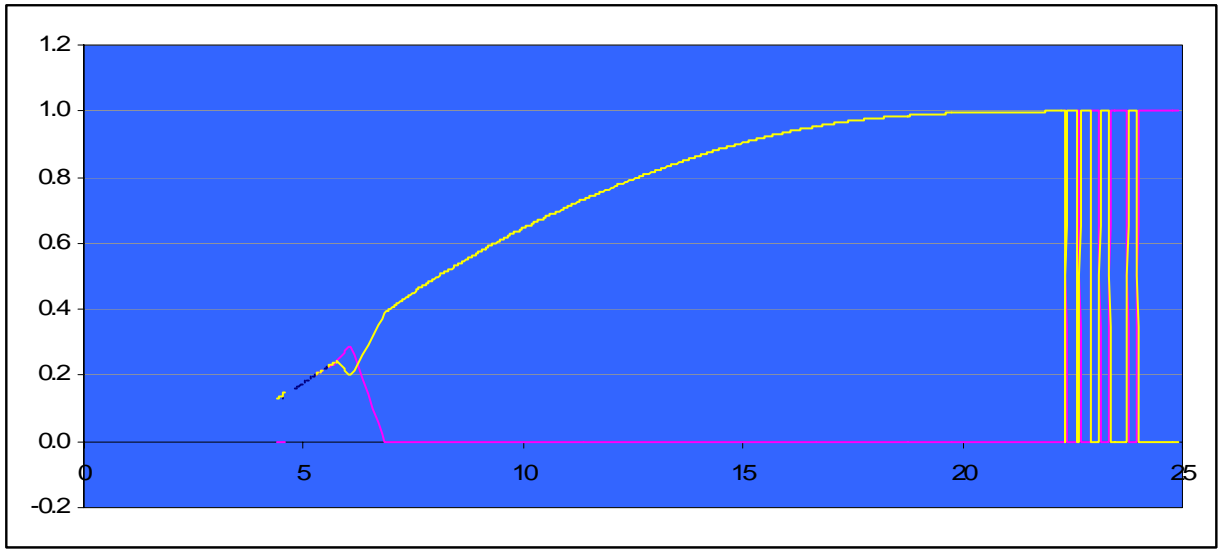

Fig.3. DIBD for map (19), γ = 0.01.

consider only the third lasing level. Having essentially longer lifetime, A*–subsystem atoms at this meta-stable level feature higher collective resistance against spontaneous emission. That means a strong restriction of their emitting activity which may be linked to reduced thermodynamic activity [12]. Using appropriate activity coefficient $\gamma^*$, one may re-write equation (5) as

(18) $\qquad\qquad\qquad\qquad\qquad\qquad\qquad\qquad\qquad\qquad \Delta G = \Delta G^0 + RT\ln[\gamma^*(1-\eta)/\eta].$

In such an approach, the active population of the A*subsystem equals to $\gamma^*[1-\eta(1-\delta_e)]$, while the ground population still is $\eta(1-\delta_e)$. Now the extended logistic map for the laser states (18) turns to

(19) $\qquad\qquad\qquad\qquad -\ln[(1-\eta)/\eta] + \ln\{\gamma^*[1-\eta(1-\delta_e)]/\eta(1-\delta_e)\} - \alpha(1-\delta_e^q) = 0.$

The activity coefficient in application to the atom emission may be roughly estimated as a ratio of the lifetimes on the uppermost level and meta-stable intermediate level, that is

(20) $\qquad\qquad\qquad\qquad\qquad\qquad\qquad\qquad\qquad\qquad\qquad\qquad \gamma^* \approx \tau_{up}/\tau_{meta}.$

An example of graphical solutions to this map with a quite modest value $\gamma^* = 10^{-2}$ is given in Fig.3. It is easy to see that even at that, definitely not the least of possible values of $\gamma^*$, the occurrence of spontaneous emission is pushed far right on the DIBD, to the area where almost all atoms belong to the A*–subsystem, making the laser functioning more stable and the thermodynamic picture more realistic. Obviously, the thermodynamic allowance is far not the only factor that permits laser to function and delays or promotes spontaneous emission, though thermodynamic prohibition is the strongest of the negative factors.

## SUMMARY

The goal of this paper was to outline a discrete thermodynamic model of laser as energy transforming device. Our basic considerations were naturally focused on 2-level laser, from which a model for 3-level device was derived using thermodynamic ideas. In real lasers, the third, lasing level helps to bypass the Einstein's prohibition by separation of the opposite impulses in time reducing the excited atoms activity.

The system of discrete thermodynamics is organized in such a way that a generally expressed external thermodynamic impact can account for any external influence on reaction (3), including, e.g., any harvesting that decreases photon or A*– populations, like non-collective irradiation stimulated by the incoming photon adsorption, and some others. Most of those effects can be rigorously formulated and incorporated into this model as thermodynamic factors.

One of the most important results of this work is immediate description of the laser's work based on the bifurcation model. Many attempts were undertaken to implement bifurcation model into lasers (e.g. [13, 14]). In our approach it happens very naturally as a solution to the basic relations. Unlike [4] that used the simplest level of the system complexity, q=1, this work is based on q=2. Bifurcation diagrams and the pumping trajectories in this case look differently, and this work results do not lead to the same conclusions. It is hard to tell at this moment all relative pros and contras of both choices of the q value.

The reader perhaps has paid attention that we have referred at some points to chemical systems. It's not occasional – in systems where "quasi-chemical" processes like A→A* and its reverse occur, or real chemical transformations, or proliferation-death processes in bio-societies, or elementary particles reactions as well, we relate the system state mostly with corresponding populations, using Boltzmann-Gibbs distribution. This feature defines uniqueness of the approaches and methods, leading to generally quite similar but very different in details results.